\documentstyle[]{l-aa}  

\begin{document}

\title{The effects of resonance scattering and dust on the UV line spectrum
of radio galaxies}

\author{
M. Villar-Mart\'\i n
\inst{1,}
\inst{2}
L. Binette
\inst{3}\thanks{The Observatoire de Lyon is a component of the
Centre de Recherche Astrophysique de Lyon: CRAL}
\and
R.A.E. Fosbury
\inst{1,}
\inst{4}
}
\offprints{M. Villar-Mart\'\i n, Dept. of Physics, University of
Sheffield, Sheffield S3~7RH, UK}

\institute{$^{1}$\,ST-ECF, Karl-Schwarschild-Str 2, D-85748 Garching,
Germany \\$^{2}$\,Max-Planck-Institute f\"ur extraterrestrische Physik,
Giessenbachstrasse, Postfach 1603, D-85740 Garching, Germany
$^{3}$\,Observatoire de Lyon, UMR CNRS 142, 9 av. Charles Andr\'e, F-69561
Saint-Genis-Laval Cedex,
France\\$^{4}$\ Affiliated to the Astrophysics Division, Space Science
Department,
European Space Agency }

\date{Received: 1995 Nov 21 ~~~ Accepted: 1996 Feb 19} 
 
\maketitle
\markboth{Resonance scattering, dust and the UV spectrum of RG}{}

\begin{abstract}

In the powerful, high redshift radio galaxies, it is believed that the
dominant source of ionization for the interstellar gas is the hard
radiation field associated with the active nucleus. The photon source
is generally external to the clouds being ionized and so the
geometrical perspective from which the gas is observed and the
presence and distribution of dust must be properly accounted for in
the diagnostic process. In this paper, we examine the formation of the
three strong lines: CIV$\lambda$1549, Ly$\alpha$\ and
CIII]$\lambda$1909 which are often observed in the nuclear and
extended emission from these sources. We find that the observed
trends, in particular the high CIV$\lambda$1549/Ly$\alpha$ ratio, are
often better explained by geometrical (viewing angle) effects than by
the presence of large quantities of dust either within or outside the
excited clouds. We show that neutral condensations along the
line-of-sight, by reflecting photons near the wavelength of
Ly$\alpha$, can increase the observed CIV/Ly$\alpha$ ratio. The
existence of HI absorption clouds (i.e., mirrors) external to the
emission region leads also to the presence of large, diffuse haloes of
what appears to be pure, narrow Ly$\alpha$\ emission.

\end{abstract}

\section{Introduction}

The strong, spatially extended, rest-frame ultraviolet emission lines
observed in high redshift radio galaxies provide one of the principal
diagnostics in establishing the state of the interstellar medium in
galaxies at early epochs.  The presence of a blue continuum and
emission lines from regions aligned with the radio axis (McCarthy
et~al. 1987; Chambers et~al. 1987) warned us that much of the observed
optical radiation might be associated with the nuclear activity and so
may not be giving us a clear picture of the stellar processes which
are of such interest in studies of galaxy formation and
evolution. Subsequent work has shown, indeed, that much of the blue
light is scattered, polarized nuclear radiation (e.g., Tadhunter
et~al. 1989; di Serego Alighieri et~al. 1993; Cimatti et~al. 1993) and
that the emission lines have a high ionization state and cannot result
from stellar photoionization (McCarthy 1993). It is clearly necessary,
therefore, to reach a clear understanding of the physical processes
involved in the formation of the various lines and continua to be able
to disentangle the stellar and the AGN-related sources.

For objects at high\,$z$, the UV rest-frame lines are shifted into the
optical band and the spectrum is generally dominated by Ly$\alpha$,
CIV$\lambda$1549, HeII$\lambda$1640 and CIII]$\lambda$1909. The
strength of the high ionization lines suggests the presence of a hard
photoionizing continuum which could originate at the AGN itself
(Robinson et~al. 1987) or be associated with fast shocks generated in
extranuclear regions by the radio jets (Sutherland, Bicknell \& Dopita
1993). The strong radio/optical asymmetries observed in these objects
which exhibit the `alignment effect' (McCarthy et~al. 1991a) may
simply result from a one-sidedness in the distribution of material. It
it is clear, however, that correlated line and continuum asymmetries
could be produced by dust scattering and line fluorescence for sources
where the radio axis falls significantly away from the plane of the
sky.

In this work, we concentrate on modeling the high excitation lines for
which rather extreme ratios relative to Ly$\alpha$ have recently been
reported. The presence of dust has been universally invoked to explain
the weakness of Ly$\alpha$ which is a resonance line and therefore,
due to multiple scattering, more susceptible to absorption. We explore
the fact that any resonance line will be extremely sensitive to
geometrical factors, an aspect of the problem which has so far been
overlooked in modeling the UV lines. If in radio-galaxies the distant
gas clouds are photoionized from the outside by partially collimated
UV radiation emitted by the nucleus, the line formation process ---
particularly for the resonance lines --- is very different from
internally ionized HII regions. The escape of resonance line photons
is strongly influenced by the presence of spaces between the line
emitting clouds.

We have collected from the literature the observed line ratios for a
number of high~z radio-galaxies in which no contribution from any
nuclear BLR is apparent. We have built a diagnostic diagram consisting
of the lines CIV$\lambda$1549/Ly$\alpha$ {\it vs.}
CIV$\lambda$1549/CIII]$\lambda$1909, in which we compare the position
of the objects with photoionization models which not only consider the
effects of internal dust but also those of the viewing perspective ---
the angle between the incoming ionizing radiation and the observer's
line of sight. Our concentration on the particular class of radio
galaxies is purely for pragmatic reasons. It is these objects, which
we presume to harbour a powerful quasar which is hidden at
optical/ultraviolet wavelengths to our line of sight, which are most
readily found and studied at the high redshifts where we have access
to the ultraviolet spectrum from groundbased observations. Our
conclusions should be equally applicable to other classes of AGN.

For some objects, Ly$\alpha$ is observed to be fainter with respect to
CIV than predicted by dust-free photoionization models. The
explanation previously proposed to explain the weakness of Ly$\alpha$
with respect H$\alpha$ or H$\beta$ has been dust destruction of
resonant Ly$\alpha$ photons. This is {\it not} borne out by our
calculations in which we have used arbitrary amounts of dust and found
that this cannot simultaneously weaken Ly$\alpha$ while leaving the
CIV/CIII] ratio relatively unchanged since resonant CIV suffers also
from dust absorption. Alternatively, by varying the proportions of the
illuminated and the shadowed cloud faces which contribute to the
observed spectrum, we are better able to match the data.  
geometric explanation could naturally explain that some of the 
brightness asymmetries noted by McCarthy et~al.  (1991a) on 
sides of the nucleus.

As we find that geometry alone (with or without internal dust) can in
principle explain most of the specific line ratios observed (fainter
Ly$\alpha$ compared with either CIV or HeII), we also discuss the
possibility of a patchy outer halo of neutral gas to account for the
diffuse Ly$\alpha$ seen in some cases to extend much beyond the CIV
emitting region and even the outermost radio lobes.  Reflection by
cold gas of the brighter Ly$\alpha$ emitting side of the ionized
clouds would lead to a narrower profile for such a diffuse
component. Another possibility is that part of the beamed nuclear {\it
continuum and BLR} radiation might be reflected at the wavelength of
Ly$\alpha$ by thin matter-bounded photoionized gas at very large
distances from the nucleus leading to a diffuse Ly$\alpha$ component
aligned with the radio axis. It appears to us that geometrical
perspective effects are an essential component of the interpretation
of the UV spectrum of radio-galaxies whether or not dust is
present. Furthermore, a spectrum in which only Ly$\alpha$ appears does
not necessarily imply starburst activity, other lines must be observed
before the existence of an HII region can be inferred.

\section{Data sample and modeling procedure}

\subsection{The data}

We have constructed a data sample containing galaxies at high~z for
which the CIV, Ly$\alpha$ and, in most cases CIII], emission lines
have been measured.  Since very high densities such as those
encountered in the BLR alter significantly the line formation and
transfer processes, we have excluded those objects which show evidence
of a BLR. In Table~1, we list the object names, the line ratios of
interest to us here, the redshift and the reference to the
observations. The larger fraction of the data are taken from the
recent thesis by van Ojik (1995) which includes objects selected on
the basis of a very steep radio spectrum. Probably by virtue of the
radio selection, these sources populate the region of the line ratio
diagram (Fig.~4) with lower CIV/CIII] ratios (lower ionization
parameter) than the previously published objects. The line
measurements refer to the integrated emission from the object
collected with a long slit aligned with the major axis.

\subsection{The model and its parameters}

The data are compared to photoionization models computed using the
multipurpose photoionization--shock code MAPPINGS~I. The version
described in Binette et~al.  (1993a,b) is particularly suited to the
problem since it considers both the effects of the observer's position
with respect to the emitting slab and the ionizing source (cf,
Fig.~3). We distinguish between the spectrum seen from the back and
from the front of the slab. The code also considers the effects of
dust mixed with the ionized gas: extinction of the ionizing continuum
and of the emission lines, scattering by the dust and heating by dust
photoionization.  The treatment of the escape of resonant CIV and
Ly$\alpha$ photons in a dusty medium is described in Appendix~B of
Binette et~al. (1993a) and is based on the results of Hummer \& Kunasz
(1980).

The dust content of the photoionized plasma is described by the
quantity $\mu$ which is the dust-to-gas ratio of the plasma expressed
in units of the solar neighbourhood dust-to-gas ratio. To specify the
gas metallicity, we scale with a factor $Z$ the solar abundance set of
trace elements from Anders \& Grevesse (1989). He/H is kept constant
at 0.1. We generally consider the solar case with $Z=1$. Since the
presence of dust implies depletion of refractory trace elements, we
use the prescription given in Appendix~A of Binette et~al. (1993a,
hereafter BWVM3) to derive the {\it gaseous phase} abundances for any
metallicity $Z$. The depletion algorithm is function of the ratio
$\mu/Z$ and makes use of the depletion indices listed in Whittet
(1992) for the different metals.

The calculations consider the gas pressure to be constant (isobaric
models) and so the density behaviour with depth in the cloud is
determined by the behaviour of the temperature and the ionization
fraction of the gas. The ionization parameter, a measure of the
excitation level of the ionized gas, is defined as the quotient of the
density of ionizing photons incident on the cloud and the gas density:

$U = \frac{\int_{\nu_o}^{\infty}{f_{\nu}d\nu/h\nu}}{c \, n_H}$

\vspace{0.2cm}

\noindent where $f_{\nu}$ is the monochromatic ionizing energy flux
impinging on the cloud, $\nu_o$ the ionization potential of H, $n_H$
the density of the gas in the front layer of the slab and $c$ the
speed of light.

\begin{table*}
\centering
\caption{Observed UV line ratios for several high~z RG with not apparent broad component }
\begin{tabular}{lllll} \hline

Name &  CIV/CIII] & CIV/Ly$\alpha$ &  $z$  & ref. \\ \hline
  Average RG &  2.054 &  0.118  &  & McCarthy (1993) \\ \hline	
MG1019+0535A & 2.12 &  1.24 & 2.76  & 	 Dey et al. 1995\\
F10214+4724 &  3.68 & 8.75 & 	2.29 & 	Elston et al. 1994	\\ 	
TX0211-122 &  3.33  &	   0.91 &  2.34 	& van Ojik et al. 1994 \\ 	
3C294   &    0.83  &      0.10     &   1.79 &  McCarthy et al. 1990a \\
0902+34   &    -- & 0.11 & 3.40 & Lilly 1988 \\
3C256-3C239 &	1.90 & 	  0.14 & 1.82 \& 1.78  & Spinrad et al. 1985	\\ 
0200+015 &  1.05 & 0.241 & 2.23	& van Ojik 1995 \\ 
0214+183 &  1.67  &	--     &   2.13 	& ~~~~~~ " \\ 
0355-037 &  1.17  &	     0.24 &  2.15 	&  ~~~~~~ " \\ 
0417-181 &  -- & 0.40 &  2.73	&  ~~~~~~ " \\ 
0448+091 & 	0.44	&     0.10  & 2.04 & 	~~~~~~ " \\
0529-549 &  0.22	 &    0.05 & 2.58 & 	~~~~~~ " \\
0748+134 &  1.29	 &    0.29 & 2.42 & 	~~~~~~ " \\
0828+193 &  0.95	 &    0.14 &  2.57 & 	~~~~~~ " \\
0857+036 &   -- & 0.39 & 2.81 & 	~~~~~~ " \\
0943-242 &  1.70	 &    0.19 & 2.92 & 	~~~~~~ " \\
1138-262 &  0.62   &     0.06 & 2.16  & 	~~~~~~ " \\
1357+007 &  -- & 0.19  & 2.67 & 	~~~~~~ " \\
1410-001 &  1.58	&     0.16 & 2.36 & 	~~~~~~ " \\
1545-234 &  0.61	 &    0.17 & 2.76 & 	~~~~~~ " \\
1558-003 &   2.25	 &    0.18 & 2.53  & 	~~~~~~ " \\
2202+128 &  -- & 0.25 & 2.71 & 	~~~~~~ " \\
2251-089 &  2.20  & -- & 1.99 & 	~~~~~~ " \\
4C23.56 &  3.40  & -- &  2.48 &	~~~~~~ " \\
4C24.28	 &   -- & 0.23 & 2.88  & 	~~~~~~ " \\
4C26.38	 &  3.71 & -- & 2.61 & 	~~~~~~ " \\
4C28.58	 & 0.17 & -- & 2.89 & 	~~~~~~ " \\
4C40.36	 &  1.05 & -- &  2.27 & 	~~~~~~ " \\
4C41.17	 &  -- & 0.05  & 3.80 & 	~~~~~~ " \\
4C48.48	 &  2.18  & -- & 2.34 & 	~~~~~~ " \\
4C60.07	 &  -- & 0.27 &  3.79 & 	~~~~~~ " \\ \hline
\hline
\end{tabular}
\end{table*}

\subsection{Adopted physical conditions}

The detailed studies of the optical emission lines of low~z radio
galaxies provide our basic reference for the properties of the
emitting gas and the ionizing continuum (e.g., Robinson et~al. 1987,
hereafter RBFT87). As a guide in choosing the input parameters, we
have assumed that the excited gas of {\it very} high~z radio galaxies
($z>2.5$) has similar properties to that of the low redshift ($z<0.1$)
objects.

For the extended (EELR) as well as the (narrow) nuclear emission
lines, photoionization by a hard continuum appears to best explain the
various line ratio diagnostic diagrams. The EELR follow a tight
sequences which are well reproduced by varying the ionization
parameter $U$ in photoionization models.  By combining
[OI]$\lambda$6300/H$\alpha$ with information on the gas excitation
(from the ratio [OIII]$\lambda 5007$/H$\beta$), RBFT87 inferred that
the ionizing spectrum is characterized by mean photon energies in the
range 30 to 40~eV, which rules out normal stars. A power law of index
$\alpha \simeq -1.4$ ($f_{\nu } \propto \nu^{+\alpha}$) obeys this
requirement and has produced remarkably good agreement with the
optical line ratios although equally good fits can be obtained using
hot ($T\simeq 150,000$\,K) blackbody spectra.  Hereafter the above
power law will be adopted as ionizing energy distribution.

Concerning metallicity, RBFT87 indicated that the abundances cannot be
much higher than solar values. They could however be lower by a factor
of a few, a distinct possibility in the case of the extranuclear gas
in very high~z galaxies. Unless specified otherwise, we have adopted
$Z=1$ in our calculations and have verified that lower abundances do
not in any way affect the conclusions reached here.

The red [SII] doublet ratio in low~z radio galaxies indicates
densities for the extended emission line regions which are not higher
than a few tens per cm$^3$.  We therefore adopt the low density
regime, specifically $n_H = 100 {\rm cm}^{-3}$, since the {\it
extranuclear} gas generally dominates the line luminosities.

\section{Models and comparison with observed UV lines}

In this section we first extend the assumption of photoionization to
modeling the line emission in the rest-frame ultraviolet of high~z
radio galaxies. We distinguish the effects of having a back and a
front view of an externally photoionized slab (section 3.2). In 3.3,
we introduce internal dust and discuss how, alone, it is insufficient
to explain the low value of the Ly$\alpha$/CIV ratio in some
objects. In 3.4 we distinguish between the resonance line `mirror'
intrinsic to our photoionized slab and the possibility of having
external cold gas --- which is shadowed from the ionizing source ---
acting as a reflector.

\input{psfig}
\begin{figure}
\par
\centerline{\psfig{figure=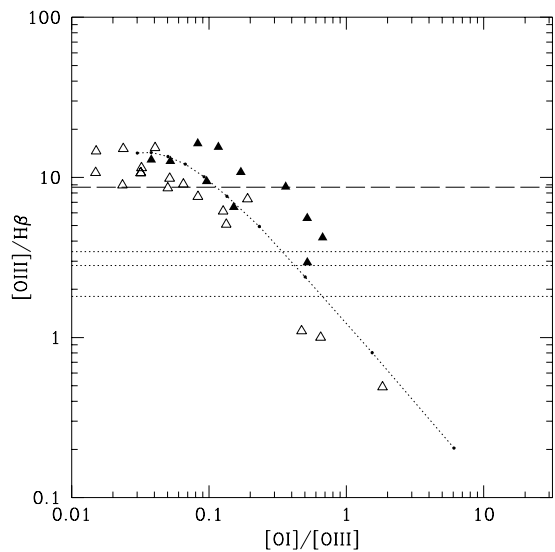,height=3in,width=3in}}
\par

\caption[]{Excitation diagnostic for the ionized gas in low~z
radio-galaxies.  Open triangles correspond to extended emission line
regions (EELR) while solid triangles correspond to nuclear regions.
The three lower horizontal lines indicate the observed [OIII]$\lambda
5007$/H$\beta$ ([OI] is not measured) ratio of very high~z
radiogalaxies (Eales \& Rawlings 1993) while the long dash line
corresponds to the ``average'' radiogalaxy spectrum as derived by
McCarthy (1993). The dotted line correspond to the sequence of
photoionization models of RBFT87.}

\end{figure}

\input{psfig}
\begin{figure}[htb]
\par
\centerline{\psfig{figure=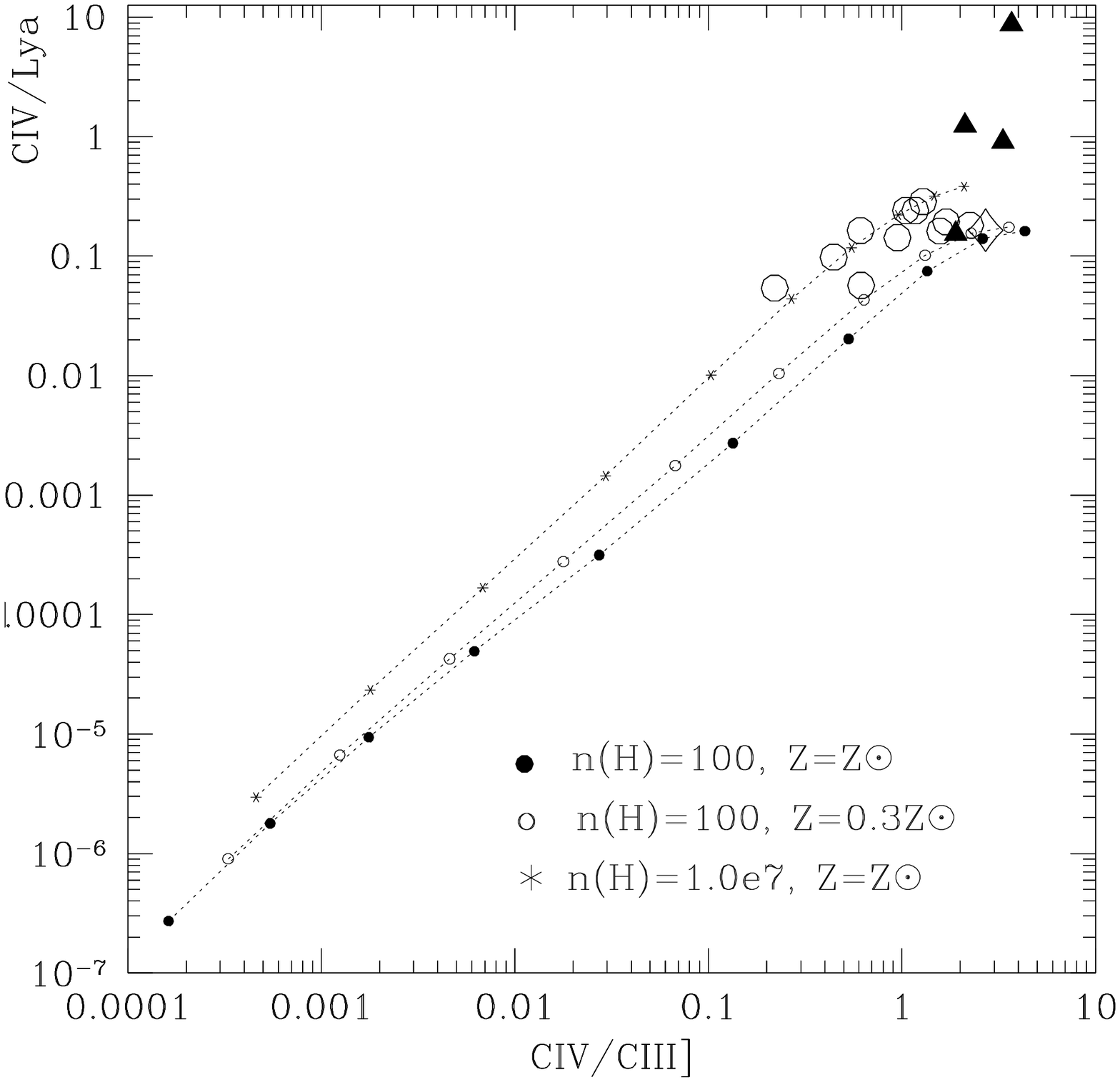,height=3in,width=3in}}
\par

\caption[]{Observed and predicted UV line ratios. Filled triangles are
the observed ratios of objects observed by different authors. Open
circles correspond to data taken from van Ojik's thesis, selected on
the basis of a very steep radio spectrum.  The open diamond is the
average radiogalaxy spectrum of McCarthy (1993).  Dotted lines
represent models in which the front and back spectra have been
summed. It is clear that high values of the ionization parameter $U$
are required to reproduce the high CIV/CIII] values observed.}

\end{figure}

\subsection{The photoionization assumption at low and high~z}

At very high redshift, observational access to the normal optical
plasma diagnostic lines is restricted and the available data set is
small. Also, where measurements are available, few lines are measured
in any given source. It is nevertheless interesting to compare the
[OIII]$\lambda$5007/H$\beta$ value observed at very high~z to that of
low~z radio galaxies as plotted in Figure~1.  The curved line
corresponds to the sequence of photoionization models of RBFT87 which
reproduce reasonably well the optical (rest frame) line ratios of the
extended emission line regions and their associated narrow nuclear
line emission. The range in $U$ is [$10^{-4}, 10^{-1}$]. The three
lower horizontal lines correspond to high~z radio galaxies while the
upper one corresponds to the ``average'' radiogalaxy spectrum as
defined by McCarthy (1993). The high~z objects are not substantially
different from the low~z sample although the apparent trend towards
weaker [OIII]/H$\beta$, if confirmed, might indicate substantially
lower metallicities: low enough to overcome the higher ratios produced
by the higher kinetic temperatures in moderately underabundant
objects. Such an effect could also be produced by insufficient spatial
resolution at high~z to separate the high from the low excitation
regions.

To see if these models reproduce the observed CIV/Ly$\alpha$ and
CIV/CIII] ratios, we have presented the observed and predicted values
in Fig.~2. The various lines correspond to calculations in which the
ionization parameter is varied, generating a sequence in $U$.
Different line types correspond to different viewing angles (front,
back and average) of the photoionized slab. The metallicity was $Z=1$
and the gas dustfree.

We find that somewhat higher values of $U$ than in the optical are
required to reproduce the CIV/CIII] ratio observed. This is most
probably a consequence of the different way in which the objects at
high and low~z are selected: the distant sources are all very powerful
radio sources with luminous AGN. For a power law of index --1.4, the
optimum value is $U\simeq 0.1$. Hereafter, diagrams will only cover
the range of [0.01,0.1] in $U$.

\subsection{Effects of viewing direction on the UV lines}
	
A notable observation is the distribution of observed points above the
model loci in Fig.~2 even at the highest value of $U$. In the extreme
objects at least --- like F~10214+4724 and TX~0211-122 --- this
results from a weakening of Ly$\alpha$ rather than atypical values of
CIV/CIII]. It is usually claimed that the destruction Ly$\alpha$
photons by resonance scattering in the presence of dust is the
explanation for its faintness, but why does {\em not} the same process
reduce CIV which is also a resonance line? Is there an alternative
explanation for this selective dimming of Ly$\alpha$?

To answer this question, we examine the geometrical aspects of the
line formation process. Each emitting cloud is approximated as a plane
parallel slab which contains a fully ionized region and a partially
ionized zone where low ionization species co-exist (e.g., O$^0$,
S$^+$, etc) with a mixture of H$^0$ and H$^+$. In principle there can
be an additional neutral zone which does not contribute to the
emission line intensities (see Fig.~3).

\input{psfig} 
\begin{figure} 
\par
\centerline{\psfig{figure=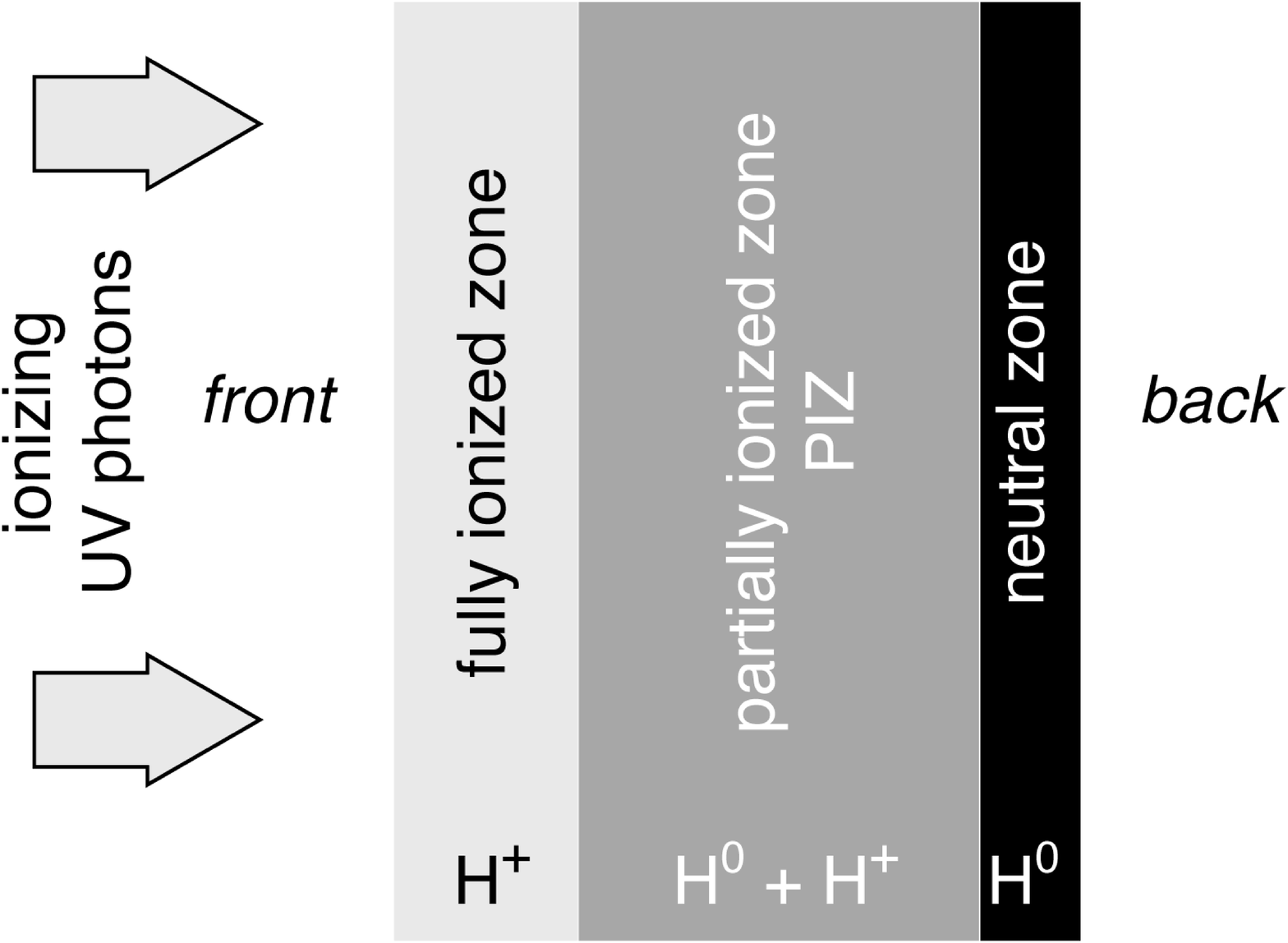,height=2.5in,width=3.5in}} 
\par
\caption[]{Adopted slab geometry for the constant pressure
photoionization calculations.The slab comprises a fully ionized zone
and a partially ionized zone (PIZ) where H$^+$ and H$^0$
coexist. Beyond the PIZ we may have a region of neutral gas.}
\end{figure}

In this section we consider the dust-free case. For most lines (like
CIII], HeII, H$\beta$, [OIII], etc) line opacity is negligible and the
line is emitted isotropically with photons escaping freely in all
directions. However, when the line opacity is important as it is for
CIV and Ly$\alpha$, line scattering occurs which increases the path
length. Another important effect of large optical depths is that the
line photon will not escape isotropically. A resonance line photon
following many scatterings must statistically escape in the direction
of highest escape probability which can be shown to be the front for
the photoionized slab depicted in Fig.~3. In the case of Ly$\alpha$,
the reason is that --- while Ly$\alpha$ photons are generated more or
less uniformly within the slab (except within the PIZ) by
recombination --- the neutral fraction and therefore the incremental
line opacity $d\tau _L/dx$ increases monotonically as a function of
depth as discussed in more detail by BWVM3. This means that for an
{\it open} geometry like that shown in Fig.~3, the zone of equal
escape probability of front {\it vs} back occurs far beyond the point
where half the luminosity of Ly$\alpha$ is produced. For the
collisionally excited CIV line, the tendency to escape from the front
also exists although it is less pronounced. It arises mainly because
the emissivity of CIV is larger towards the front due to the
temperature gradient across the C$^{+3}$ zone. Note that while CIV is
emitted and scatters within the rather limited zone containing
C$^{+3}$, Ly$\alpha$ remains subject to scattering outside the region
where it is produced. While most Ly$\alpha$ emission is produced in
the fully ionized zone, most of the line opacity occurs within the
PIZ. The presence of a layer of neutral gas beyond the PIZ will
increase the anisotropy of Ly$\alpha$ escape.

\input{psfig}
\begin{figure}
\par
\centerline{\psfig{figure=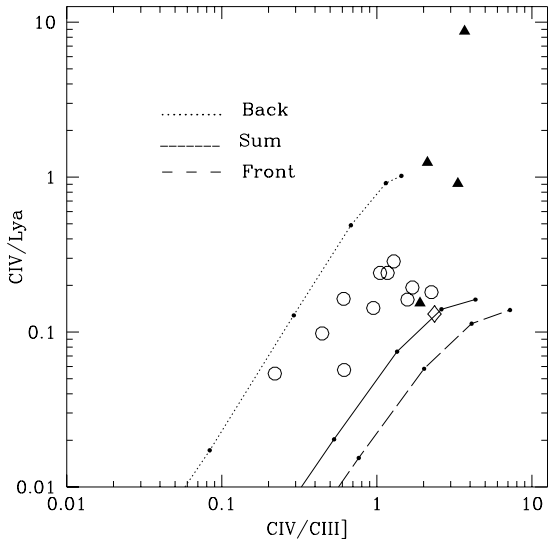,height=3in,width=3in}}
\par

\caption[]{Influence of viewing geometry on the UV line ratios. The
dotted line corresponds to the line spectrum seen from the {\it back}
of the slab (cf, Fig~3) while the dashed line corresponds to the
spectrum seen from the {\it front} (UV irradiated face). It is
apparent that perspective plays a very important role on the
CIV/Ly$\alpha$ ratio. The solid line represents models obtained by
summing the back and front spectra which would represent symmetric
case where equal numbers of clouds are observed with shadowed and
illuminated faces. }

\end{figure}

The effects described qualitatively above are shown in Fig.~4 using
detailed photoionization calculations. We present the same sequence of
dust-free models as in Fig.~2 but distinguish between the spectrum
seen from the {\it back} --- equivalent to observing the clouds
through the PIZ --- from that seen from the {\it front} --- equivalent
to seeing the UV irradiated side. The differences are striking: both
Ly$\alpha$ and to a lesser extent CIV are fainter when seen from the
back. The CIII] line is isotropic in the dust-free case. The fact that
Ly$\alpha$ is more affected by perspective is due to the significant
amount of neutral hydrogen (i.e., large line opacity) in the PIZ which
acts as a mirror.  Although we have considered a very simplified
geometry in our calculations, the method nevertheless treats properly
the essential physical effects and indicates how important the viewing
direction is in this open geometry.

Fig.~4 suggests that perspective effects alone ({\it without any
dust}) are sufficient to explain the weak Ly$\alpha$ seen in some
objects. Ionization bounded calculations with $U=0.1$ imply total
hydrogen column densities (H$^+$ region + PIZ) $N_H \sim 10^{22}
\,{\rm cm}^{-2}$ (of which about 60\% is ionized).  Adding a modest
neutral zone beyond the PIZ of $\simeq 4\, 10^{21} \,{\rm cm}^{-2}$
would double the CIV/Ly$\alpha$ ratio without affecting in any way the
CIV/CIII] ratio. It seems, therefore, that a geometry where we see
preferentially the ionized gas from the side of the PIZ gives us an
explanation of the weakness of Ly$\alpha$.

How would this apply to the EELR of the powerful radio galaxies? In a
very simplified scheme, we can imagine (see Fig.~5) that the clouds
seen from the nearside cone are seen from a direction which we
approximate as the back perspective in our slab calculations while
clouds on the far side would be seen from the front. The studies of
McCarthy et~al. (1991a) which emphasized the one-sideness of the line
brightness suggest that the observer with limited spatial resolution
at very high~z will be biased towards either a back or front dominated
perspective depending on whether it is the near or the farside
illuminated cone which is intrinsically brighter. Our proposed
interpretation of the high CIV/Ly$\alpha$ ratio of some objects in
Fig.~4 is that they correspond to the case where the brightest clouds
are seen from behind. Note that, because of the intensity weighting,
there is rather little difference between the pure front perspective
and the sum of equal numbers of front- and back-clouds, a small effect
which is not easily distinguished from that of slightly reducing the
value of $U$.

\input{psfig}
\begin{figure}
\par
\centerline{\psfig{figure=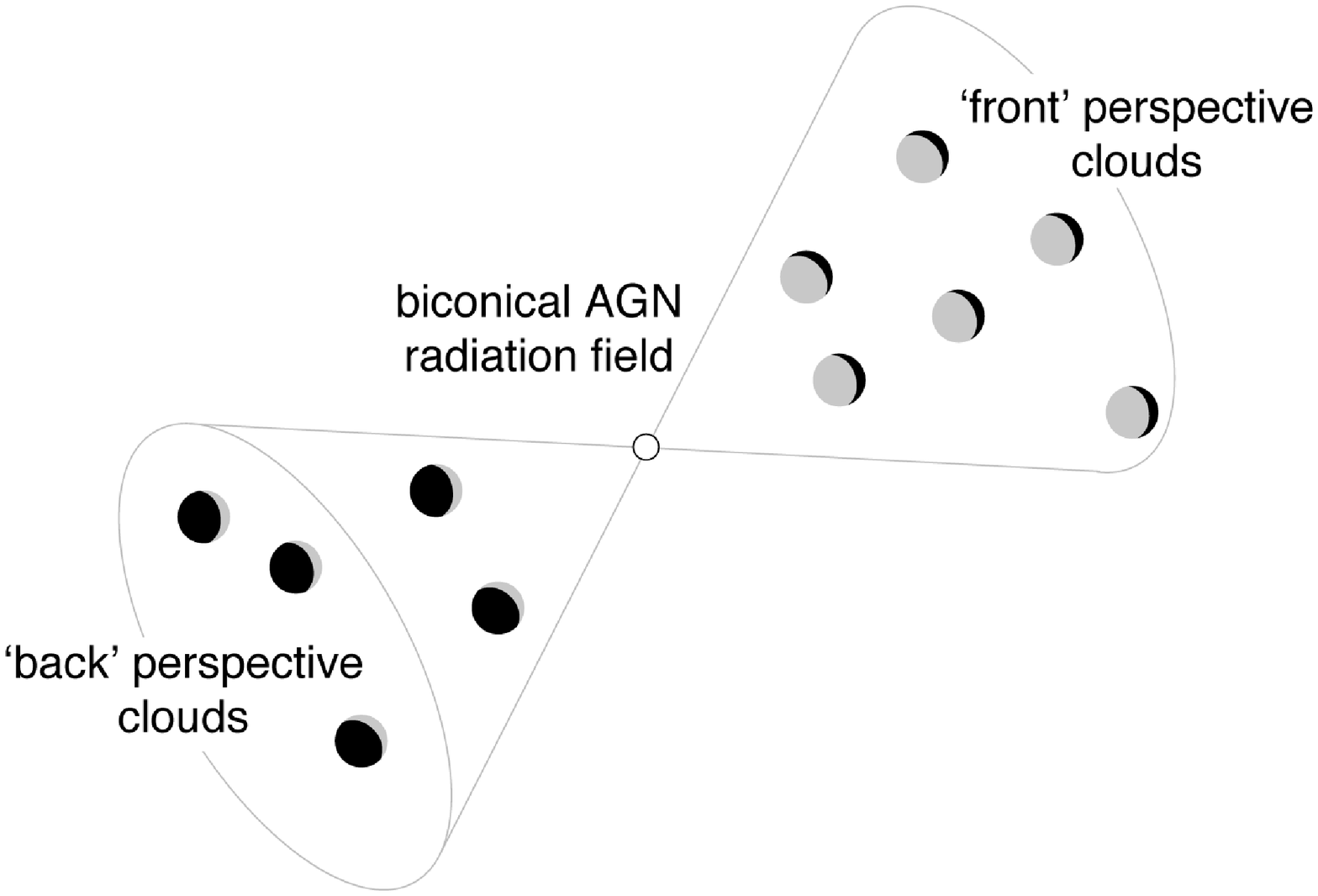,height=2in,width=3.2in}}
\par

\caption[]{}

\end{figure}

Our conclusion from this section is that perspective effects in an
open --- externally illuminated cloud --- geometry go a long way
towards explaining the behaviour of the CIV, Ly$\alpha$, CIII] line
ratio diagram.  It implies that, when spatially resolved spectra
become available, marked asymmetries in the CIV/Ly$\alpha$ ratio could
arise when the axis of the illuminated cones forms a large angle with
the sky plane.

\subsection {The effects of internal dust}

There is evidence for the existence of dust in some high~z radio
galaxies. The detection of 4C41.17 ($z=3.8$) and B2~0902+34 ($z=3.5$)
(Chini \& Kr\"ugel 1994; Dunlop et al. 1994) and 8C1435+635 ($z=4.26$)
(Ivison 1995) in the mm spectral range is attributed to warm
dust. Also, the IRAS galaxy F10214+4724 ($z=2.29$) has been shown,
from the far infrared flux, to contain $\sim 10^8 M_{sol}$ of dust,
although this figure may be reduced by the gravitational lensing
amplification factor (Serjeant et~al. 1995). In addition, the mid- to
far-infrared measurements at intermediate~z are explained as emission
from warm dust (Heckman, Chambers \& Postman 1992). We should
emphasize, however, that we have no direct measurement of how this
dust is spatially distributed. In the event that this infrared
emission arises from the reprocessing of higher energy photons from
the AGN by a dusty torus, it has no direct bearing on our modeling of
ionized gas at tens of kpc. However, there is considerable evidence
that aligned blue polarized continuum is the result of scattering of
the anisotropic nuclear radiation field by dust (e.g., Cimatti
et~al. 1993). In this case it is very likely that at least some of the
dust is internal to the extended line emission regions. In this
section we consider the effects of including dust within the gas
clouds which emit the UV lines.

\input{psfig}
\begin{figure}
\par
\centerline{\psfig{figure=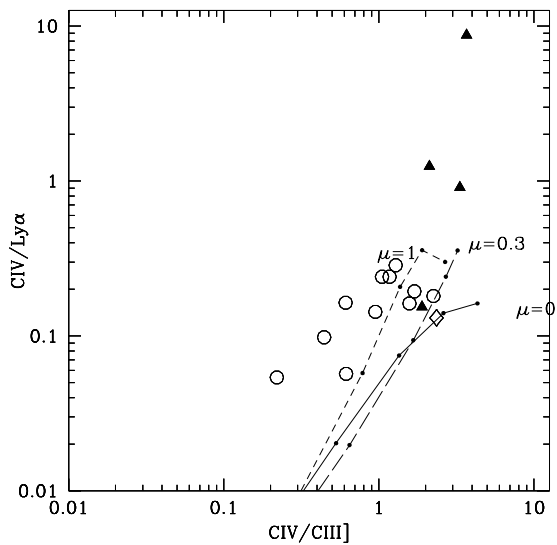,height=3in,width=3in}}
\par

\caption[]{Effects of varying the amount of internal dust as seen from
the front perspective. Short dashed line corresponds to models with
$\mu=1.0$, the long-dashed to models with $\mu=0.3$ and solid line to
dust free models.}

\end{figure}

We illustrate the effects of dust mixed with the emitting gas in
Fig.~6 where we plot sequence of models which correspond to the front
perspective for three different dust-to-gas ratios: $\mu=0$, 0.3 and
1. The effects on the line ratios are evident. The resonance
scattering suffered by CIV and Ly$\alpha$ increases their pathlengths
many times and, therefore, the probability of their being absorbed by
dust grains is much higher than for CIII]. In the case of Ly$\alpha$,
however, the geometrical thickness of the H$^+$ region exceeds greatly
that of the C$^{+3}$ since we observe more than one stage of
ionization of metals (e.g. C$^{+2}$). Any reasonable parameters for
the ionization structure of a photoionized slab with $Z\sim 1$
indicates that the opacity in Ly$\alpha$ greatly exceeds that of CIV,
which implies a larger pathlength increase for Ly$\alpha$ than for
CIV. This results in relatively more Ly$\alpha$ absorption by dust.
This effect explains how the ratio CIV/Ly$\alpha$ increases somewhat
with increasing $\mu$.  Dust absorption of resonant CIV on the other
hand causes a comparable decrease in CIV/CIII]. What is important in
these results is that, even with a concentration of internal dust as
high as $\mu =1$ (equivalent to that in solar neighbourhood cold
clouds), it is not possible to reproduce the high ratio CIV/Ly$\alpha
\ge 1$ which is observed in some objects and has been attributed to
dust. We emphasize that higher amounts of dust do not change these
results. For instance models with $Z=\mu=2$ do not lead to any higher
Ly$\alpha$/CIV ratio than the models shown in Fig.~6.

\input{psfig}
\begin{figure}
\par
\centerline{\psfig{figure=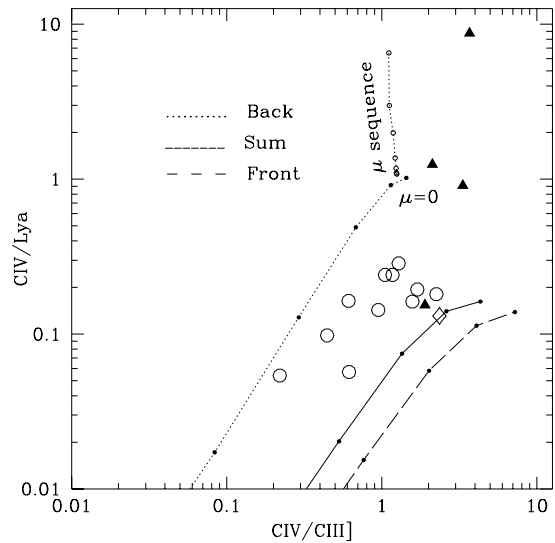,height=3in,width=3in}}
\par

\caption[]{Effects of perspective combined with small quantities of
internal dust. The dust-free $U$ sequences of Fig.~5 are repeated in
this figure. The nearly vertical dotted line with open circles
corresponds to the last model with $U=0.1$ (but with metallicity $Z=
0.3$) in which the dust content is increased in proportion up to
$\mu=0.017$. Larger quantities of dust do not produce any further
increase in Ly$\alpha$/CIV as seen from the back since it severely
extinguishes both the CIII] and CIV lines which are produced in the
front layers. }

\end{figure}

When the clouds are viewed from behind, even with $\mu =0.3$ they are
sufficiently opaque that all the UV lines are severely absorbed. With
$\tau_V = 5 \, 10^{-22} \mu N_H \sim 5$ and $\mu =1$, all of the high
excitation UV lines become absorbed within the PIZ. To produce an
acceptable spectrum, without reddened CIV/CIII] and CIV/HeII ratios,
as seen from the back of an ionization bounded slab with $U=0.1$, we
have to use much smaller amounts of dust like $\mu \simeq 0.017$ (2\%
of local ISM) in models. In Fig.~7, we plot the back and front
sequences for such models. These can reproduce the weak Ly$\alpha$
objects although this result is obtained only for the back spectrum,
emphasizing that perspective is the dominant factor. It should be
noted that the amount of extinction within the ionization bounded slab
implied by $\mu = 0.017$ (i.e., $A_V \simeq 0.1$--0.2) is consistent
with the quantity of small dust grains needed to explain the extremely
blue continuum of the ``detached'' ionized cloud in PKS 2152-69 (di
Serego Alighieri et~al. 1988; Magris \& Binette private
communication), supposing that the continuum energy distribution is
the result of dust scattering of the nuclear radiation.

Note that because the scattering/absorbing dust is locally
kinematically linked to the emission line gas, we require much smaller
column densities of HI than the absorbing screen proposed by van~Ojik
et~al. (1994) ($\sim 10^{23} \,{\rm cm}^{-2}$) to explain
CIV/Ly$\alpha \sim 1$. The line widths and centroids of the PIZ and of
the fully ionized gas are expected to be quite similar within each of
the emitting clouds, the ensemble of which could have a greater
`turbulent' velocity dispersion.

To reproduce the extreme case of the IRAS galaxy F10214+4724 with
CIV/Ly$\alpha \sim 10$, we need additional neutral gas beyond the
PIZ. For instance, it requires only a column density of $N_{H^0} \sim
1.5 \, 10^{21} \,{\rm cm}^{-2}$ assuming $\mu=0.017$. Alternatively we
might consider that $\mu$ within the PIZ increases with depth as the
degree of ionization decreases. In this case, no additional HI gas
would be required. If this IRAS galaxy is indeed an extreme Seyfert~2
as argued by Elston et~al. (1994), then a closed, dust enshrouded
geometry as proposed by BWVM3 to explain the Lyman and Balmer
decrements in Seyfert~2 would be more appropriate than the open
geometry adopted here which may be applicable only to the truly
extended large scale gas. It is likely that objects with unusually
strong NV$\lambda 1240$ emission (such as in F10214+4724 and
TX0211-122) are cases where the NV originates predominantly from the
inner NLR. A high NV/CIV ratio indicates very enriched gas which is
not unexpected within the inner parts of an AGN (Hamman \& Ferland
1993) and is consistent with the lack of convincing evidence that NV
is spatially resolved in any of these objects.

\subsection{Neutral gas mirrors}

So far we have considered the effects of scattering by gas which forms
part of the line emitting clouds: we refer to this as an `intrinsic'
process. Similar effects could be produced by a large-scale
distribution of predominantly neutral material surrounding the
emitting regions: we will refer to this as the `extrinsic' case. Any
extrinsic neutral gas component with a non-negligible covering factor
could affect the observed spectra in a way which mimics that of the
back perspective described earlier. Let us suppose that this outer
material is broken up into cold gas clumps which are randomly
distributed. In such a case, some Ly$\alpha$ photons which leave the
ionized cones will escape the region through the holes between the
external clumps while others will strike the neutral clumps and be
immediately scattered away to escape eventually through another hole
in a different direction (see Fig.~8). If observed with sufficient
spatial resolution, such a geometry would result in holes in the
Ly$\alpha$ brightness due to reflection by intervening clumps as well
as diffuse sources corresponding to reflection from clumps on the far
side of the source. The bulk of the Ly$\alpha$ luminosity would be
preserved but redistributed on an apparently larger scale than the
true line emitting clouds.  Only a more closed geometry would result
in a significant destruction of Ly$\alpha$, assuming that the
interclump space does not contain pure dust segregated from the gas
phase.

The reflection efficiency of clouds will, of course, depend strongly
on the relative velocity fields of the emitting and the cold
regions. If the extended gas has a large scale ordered motion but
small `microturbulence', the reflection effects would be
localized. Broad Ly$\alpha$ emission and continuum from the AGN could,
however, be scattered by any of the extranuclear clouds (see
Sect.~3.4.2). In general, we would expect the diffuse, scattered
Ly$\alpha$ to show a narrower line than the integrated emission
profile.

We will review the observational evidence for the existence of such
large scale mirrors.

\subsubsection{The radio galaxy PKS2104-242}

PKS2104-242 (McCarthy et~al. 1990b) shows very extended emission lines
of Ly$\alpha$, CIV and HeII associated with continuum knots which lie
between the radio lobes and are aligned with the radio axis. The
Ly$\alpha$ image shows emission resolved into three distinct clumps,
two of them corresponding roughly with the two continuum knots. In
addition, there appears to be a low surface brightness halo of
emission surrounding the entire object. Is this halo a consequence of
reflection by neutral material of either Ly$\alpha$ emission or of the
intense nuclear {\it continuum}, or are these photons emitted locally
by H$^+$ recombination? A way of discriminating between these two
possibilities is to look for the detection of any other emission line
in the halo. If Ly$\alpha$ is the result of reflection by cold HI,
there will be {\it no} other lines (except possibly resonant MgII). If
it is instead produced by recombination, other lines should be
detected in the rest-frame optical band like H$\alpha$,
[OIII]$\lambda$5007, etc.

\input{psfig}
\begin{figure}
\par
\centerline{\psfig{figure=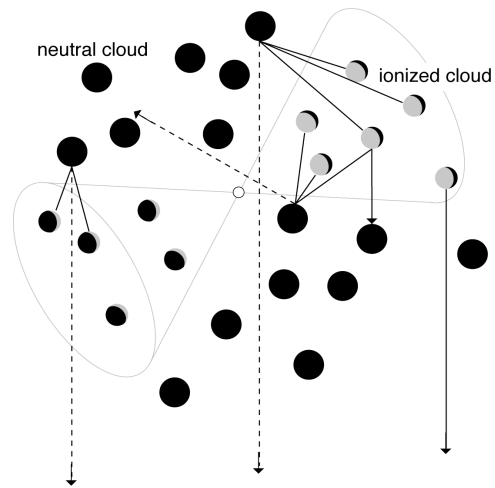,height=2.5in,width=2.5in}}
\par

\caption[]{The extrinsic case. Neutral material external to the
ionized regions can strongly influence the appearance of the object in
Ly$\alpha$. In this figure, the neutral clumps (black clouds) cover a
significant fraction of the ionized regions (grey+black clouds). The solid
arrows represent Ly$\alpha$ photons emitted directly in the ionized
regions. Dashed arrows represent Ly$\alpha$ photons that after striking
neutral clumps are reflected in other directions. The reflection by
the neutral H atoms is so effective that the dust has little chance to
interact with the photons before they find a hole to escape.
}

\end{figure}

\subsubsection{The radiogalaxy 3C~294}

From images of this object, McCarthy et~al. (1990a) report the
existence of Ly$\alpha$ emission extending over a region covering
170kpc. The Ly$\alpha$ emission is elongated and well aligned with the
inner radio-source. An intriguing aspect of the observations is the
one-sidedness of the CIV emission (obtained with an long slit aligned
with the axis of elongation). The decreasing linewidth of Ly$\alpha$
on the side where CIV is absent (South) suggests to us the possibility
that Ly$\alpha$ in the south corresponds to scattered {\it continuum
and BLR Ly$\alpha$ photons} by HI clumps lying along this direction
provided the ionizing radiation has already been reprocessed (filtered
out) within the nuclear regions into NLR or BLR emission.

Failing that (scattering by HI clumps), an alternative possibility
resides in very thin matter-bounded photoionized sheets of gas which
can be very efficient at reflecting the intense (beamed) nuclear
continuum as well as BLR Ly$\alpha$ photons. Due to its large
scattering cross section, very small column depths of H$^0$ (a trace
specie within the ionized phase) are sufficient to scatter effectively
the impinging flux within the core of the Ly$\alpha$ thermal profile.
If we suppose the existence within the cone of ionization of the
radio-galaxy of a population of clouds of similar physical conditions
to that thought to apply to Ly$\alpha$ forest clouds, the energy
reflected due to resonant scattering by HI typically exceeds that
generated within the clouds by reprocessing of the ionizing radiation
(i.e. by recombination). In effect, adopting $N_{H^0} = 10^{13.8}
\,{\rm cm}^{-2}$ as a typical column density of such cloud, we derive
an equivalent width in absorption (i.e. the scattered intensity) of

$$EW^{abs}_{Ly\alpha} \approx 0.15 \sqrt{{\rm ln}(4.2\,10^{-14}
N_{H^0})} \,{\rm \AA }=0.15{\rm \AA } $$

\noindent which corresponds to the saturated part of the Ly$\alpha$
curve of growth, assuming $b=18.2$\,km\,s$^{-1}$. Adopting for
definiteness $U=0.1$ and $\alpha =-1.4$, we obtain that the fraction
of reprocessed ionizing photons, $F_{MB}$, within a very thin
photoionized sheet\footnote{With these parameters, the total column
density is given by $ N_{H} = N_{H^+} \approx 2.3 \, 10^4 N_{H^0}$.}
is given by

$$F_{MB}\approx 1.8 \, 10^{-18} N_{H^0}= 1.1 \, 10^{-4}$$

\noindent The equivalent width of Ly$\alpha$ in emission as a result of
recombination is simply

$$EW^{em}_{Ly\alpha} \approx 390\, F_{MB} \, C_f \, {\rm \AA}=0.04\, C_f \,{\rm \AA}$$

\noindent where $C_f$ is the covering factor. If we integrate the
absorbed intensity over the same area as that given by $C_f$, the
ratio of scattered versus emitted Ly$\alpha$ is $\sim 4$ (the ratio
between the above two equivalent widths). Allowing for the fact that
the cloud will be exposed not only to the nuclear continuum but to the
BLR Ly$\alpha$ which typically peaks at twice the continuum level, the
ratio can be expected to increase up to 8 times the Ly$\alpha$ emitted
by recombination. To account for the observed FWHM $\la 1000$\,
km\,s$^{-1}$ in 3C294 (South), one most rely on a turbulent velocity
field for the the Ly$\alpha$ clouds.

We ought to consider seriously the possibility that Ly$\alpha$ on the
southern knot of 3C294 corresponds to large scale scattered light by
either HI or even ionized gas unless other bona fide emission lines
were detected. So far, no other lines than Ly$\alpha$ are observed
which is consistent with our suggestion. The only certain way of
discriminating between reflection and {\it in situ} emission would be
the detection at this radius of {\it any} non-resonance line in either
the optical ([OII], [OIII], H$\beta$ etc.) or the UV (CII], CIII]). To
conclude, we believe that HII regions or starbursts should not be
considered the default emission mechanism in cases where only
Ly$\alpha$ is present, especially along the axis where the AGN
supposedly collimates its intense nuclear continuum~+~Ly$\alpha$(BLR)
light.

\subsubsection{Ly$\alpha$ absorption in 0943-242}
	
The radio galaxy 0943-242 shows a Ly$\alpha$ profile with several
absorption features which cover the whole of spatial extent of the
Ly$\alpha$ emission (R\"ottgering et~al. 1995). The strongest of the
absorbers is at least as extended spatially as the emitting region
($\sim 13$\,kpc) and its HI column density is $10^{19} \,{\rm cm}^{-2}$. The
absorption line is blue-shifted by 250km~s$^{-1}$ with respect to the
emission peak. The `screen' is kinematically distinct from the
emitting gas and therefore clearly external to the emitting
clouds). This is a clear demonstration that HI screens --- or mirrors
depending on the perspective --- exist on galaxy scales at these early
epochs. From Fig.~5 of R\"ottgering et~al., we estimate that this
cloud would reflect $\sim 30$\% of the Ly$\alpha$. Thicker clouds
could exist around other objects but would be hard to detect when only
the faint wings at the extremities of the emission profile were
transmitted. In the case of 0943-242, even if the screen contained
dust, its effects would be negligible. The reason, as stated before,
is that Ly$\alpha$ photons (seen by the cloud) are incident from the
exterior and will be very effectively scattered away before being
absorbed by the dust.  Even the photons getting through (i.e., without
any scattering) to us because they are sufficiently far in the wings
of the profile would not see much dust in this particular cloud. With
such low column densities, $\tau_V \sim 5 \, 10^{-22} \mu N_H =
7.5\,10^{-5}$ for $\mu=0.015$ and 0.005 for $\mu=1$. Much thicker
clouds may result in some non-negligible extinction but again this
does not arise because of the multiple scattering of Ly$\alpha$.

\section{Conclusions}

If the large-scale extended emission line regions (EELR) in radio
galaxies are photoionized predominantly by the collimated UV radiation
emitted by a hidden AGN, the line emission and transfer processes are
characterized by what we call an 'open' geometry with externally
illuminated clouds. This produces an emission line spectrum which is
distinctly different from an internally ionized HII region, especially
in the ultraviolet.

Our principal conclusion is that internal dust, even in proportions as
large as that found in the solar neighbourhood, does not
satisfactorily explain the large CIV/Ly$\alpha$ ratios commonly seen
in groundbased observations of high redshift radio galaxies. Note that
the maximum amount of dust considered here ($\mu=1.0=Z$) should be
considered a sensible upper limit on the ground that: a) the extended
large scale gas ($\gg 5$kpc) in radio galaxies is likely to be less
than solar in metallicity, b) the physical conditions found within the
higly ionized warm plasma threatens dust survival and a more realistic
approach would favour using $\mu \ll Z$. In any event, more dust does
not lead to any higher Ly$\alpha$/CIV ratio as we have verified with
models in which $Z=\mu=2$.

We have demonstrated that geometrical effects can, in most cases
without invoking any dust, explain the observed trends in the
data. Although we agree that internal dust can quite effectively kill
the Ly$\alpha$ luminosity, we consider essential to examine
also its effects on the other UV lines --- especially CIV which is
also a resonance transition. 

Although some internal dust within the ionized gas is allowed by our
modeling and even improve things in a few cases, we find that the
effects of geometrical perspective {\it dominate} the intended
decrease of Ly$\alpha$ without affecting excessively the other lines.
The current work demonstrates that the faintness of Ly$\alpha$ is not
sufficient per~se to imply the existence of large amount of dust in
these objects.

The influence of the geometrical factors discussed above on the line
spectrum gives us, potentially, a natural explanation for some of the
emission line asymmetries seen in the 'alignment effect' in radio
galaxies.

We have also investigated the effects that material external to the
ionized cones will have on the appearance of the object in the light
of Ly$\alpha$. We suggest that the existence of the diffuse Ly$\alpha$
halos observed in some high~z radio galaxies could be due to the
reflection of Ly$\alpha$ photons by neutral clumps lying outside the
ionization cones. In addition to reflecting Ly$\alpha$, such neutral
clouds could be responsible for the spatially extended absorption seen
within the Ly$\alpha$ emission profile. Pure Ly$\alpha$ emission
clouds need not, therefore, necessarily be identified as star-forming
regions.

\begin{acknowledgements} MV-M acknowledges support from the Deustche
Forschungsgemeinschaft; also thanks the Observatoire de Lyon for its
hospitality during a collaborative trip in June 1995 and Jacco van
Loon for useful discussions.  
\end{acknowledgements}

\end{document}